\documentclass[10pt,superscriptaddress,aps,pra,twocolumn,longbibliography]{revtex4-2}
\usepackage{amsmath,amssymb,amsthm}
\usepackage{multirow}
\usepackage{tabularx}
\usepackage[breaklinks=true,
colorlinks,
citecolor=magenta,
linkcolor=magenta,
urlcolor=magenta]{hyperref}
\usepackage[pdftex]{graphicx}
\usepackage{times,txfonts}
\usepackage{orcidlink}
\usepackage{braket}
\usepackage{color}
\usepackage{natbib}
\usepackage{url}
\usepackage{xurl}
\usepackage{makecell}
\usepackage{bm}
\usepackage{amsmath,blkarray}
\usepackage{mathtools}
\usepackage{makecell}
\usepackage[normalem]{ulem}
\usepackage{latexsym}
\usepackage{tabularx, booktabs}
\usepackage{graphics,epstopdf}
\usepackage{graphicx}
\usepackage[linesnumbered,lined,ruled,vlined]{algorithm2e}
\usepackage{amsfonts}
\usepackage{tikz}
\usepackage{graphicx}
\usepackage{subfigure} 

\usepackage{graphicx}
\usepackage{dcolumn}
\usepackage{bm}

\begin{document}

\title{Programmable Coherent Memory Kernels for Quantum Reservoir Computing in Waveguide QED}

\author{Hong Jiang\orcidlink{0009-0008-7937-6671}}
%\email{jiang.hong@iff.csic.es}
\affiliation{Quantum Advanced Research Center, QuARC-CSIC, Calle Serrano 113b, 28006 Madrid, Spain}
\affiliation{Instituto de Física Fundamental, IFF-CSIC, Calle Serrano 113b, 28006 Madrid, Spain}

\author{Yu Wu\orcidlink{0009-0002-3998-3342}}
\affiliation{Department of Physics, Shanghai University, 200444 Shanghai, China}
\affiliation{Instituto de Ciencia de Materiales de Madrid ICMM-CSIC, Cantoblanco, 28049 Madrid, Spain}

\author{Yue Ban\orcidlink{0000-0003-1764-4470}}
\email{yue.ban@csic.es}
\affiliation{Quantum Advanced Research Center, QuARC-CSIC, Calle Serrano 113b, 28006 Madrid, Spain}
\affiliation{Instituto de Ciencia de Materiales de Madrid ICMM-CSIC, Cantoblanco, 28049 Madrid, Spain}

\author{Xi Chen\orcidlink{0000-0003-4221-4288}}
\email{xi.chen@csic.es}
\affiliation{Quantum Advanced Research Center, QuARC-CSIC, Calle Serrano 113b, 28006 Madrid, Spain}
\affiliation{Instituto de Ciencia de Materiales de Madrid ICMM-CSIC, Cantoblanco, 28049 Madrid, Spain}

\author{Juan Jos\'e Garc\'ia-Ripoll\orcidlink{0000-0001-8993-4624}}
\email{jj.garcia.ripoll@csic.es}
\affiliation{Quantum Advanced Research Center, QuARC-CSIC, Calle Serrano 113b, 28006 Madrid, Spain}
\affiliation{Instituto de Física Fundamental, IFF-CSIC, Calle Serrano 113b, 28006 Madrid, Spain}

\date{\today}
\begin{abstract}
Reservoir computing requires
not only
long memory,
but also
memory distributed over the timescales relevant to a given task. 
In most quantum reservoir computing architectures, however, the memory profile emerges from intrinsic dynamics and cannot be directly engineered.
Here we introduce a waveguide-QED quantum reservoir in which 
coherent memory return is controlled by the coupling geometry.  
The reservoir consists of a driven Kerr-nonlinear resonator coupled to a one-dimensional waveguide at spatially separated points. An unmeasured propagating field carries information about past inputs and coherently returns it to the resonator after a finite propagation time, forming a non-Markovian feedback channel without intermediate measurement or classical reinjection.
Delay-resolved memory benchmarks reveal two complementary functions: waveguide propagation determines when past information returns, whereas Kerr dynamics convert the returned field into nonlinear temporal features accessible through a linear readout. NARMA-$n$ prediction shows that the optimal feedback delay shifts systematically with task order and approximately tracks the explicit input-product lag. These results establish coherent delayed feedback as a mechanism for engineering task-relevant memory in quantum reservoir computing.

\end{abstract}

\maketitle

\section{Introduction}
\label{sec:introduction}

Reservoir computing (RC) processes temporal information
by embedding
input histories into the high-dimensional dynamical state of a complex system, while training only a linear readout layer~\cite{jaeger2001,maass2002,Appeltant2011,larger2017,tanaka2019,hulser2022,zhang2023,liang2024,ren2024,iacob2024,lugnan2025}.
Building on this framework, quantum reservoir computing (QRC) exploits coherence, intrinsic interactions, and large accessible quantum state spaces to generate rich temporal features,  motivating extensive theoretical proposals and a growing number of experimental demonstrations across photonic,
superconducting-circuit, trapped-ion, spin-based, and related
platforms~\cite{govia2021,Mujal2021,kalfus2022,spagnolo2022,bravo2022,senanian2024,hou2026,zhu2025,zhu2025b,das2026,prieto-garcia2026,kobayashi2026,paparelle2026,zhang2026}. Despite these advances, realizing the full computational potential of QRC requires not only rich quantum dynamics, but also the ability to retain and transform information from past inputs.

Memory is a central resource in RC: a reservoir must preserve task-relevant information from past inputs  while progressively discarding obsolete history~\cite{cucchi2022,mujal2023,kobayashi2024,sannia2026,vrugt2026}.
However, the usefulness of memory is not determined by duration alone. Standard temporal tasks, including short-term memory (STM)~\cite{fette2005}, delayed-product~\cite{sannia2025}, delayed-XOR~\cite{iacob2024a} and NARMA-$n$ prediction~\cite{vrugt2026}, require information retained at specific past lags or over characteristic temporal windows. The relevant computational resource is therefore not simply memory depth, but a delay-resolved memory profile aligned with the temporal structure of the task. In QRC, quantum coherence and intrinsic interactions offer distinct physical mechanisms for retaining past information and transforming it into nonlinear temporal features accessible to the readout~\cite{mujal2023,kobayashi2024,sannia2026}.

This task-matching perspective has been explored extensively in classical RC through 
task-specific capacity analysis~\cite{hulser2023} and delay-based architectures, where self-feedback loops or inter-node propagation delays reshape
lag-resolved memory profiles and information-processing capacities~\cite{donati2024,
iacob2024,iacob2024a,mullarkey2025}.
The coherent quantum counterpart, however, remains less developed. 
In a time-delay quantum reservoir,
past information must be carried by an unmeasured propagating field and returned coherently to the reservoir, rather than being converted into a classical record and subsequently reinjected. 

\begin{figure}[t]
    \centering
    \includegraphics[width=\linewidth]{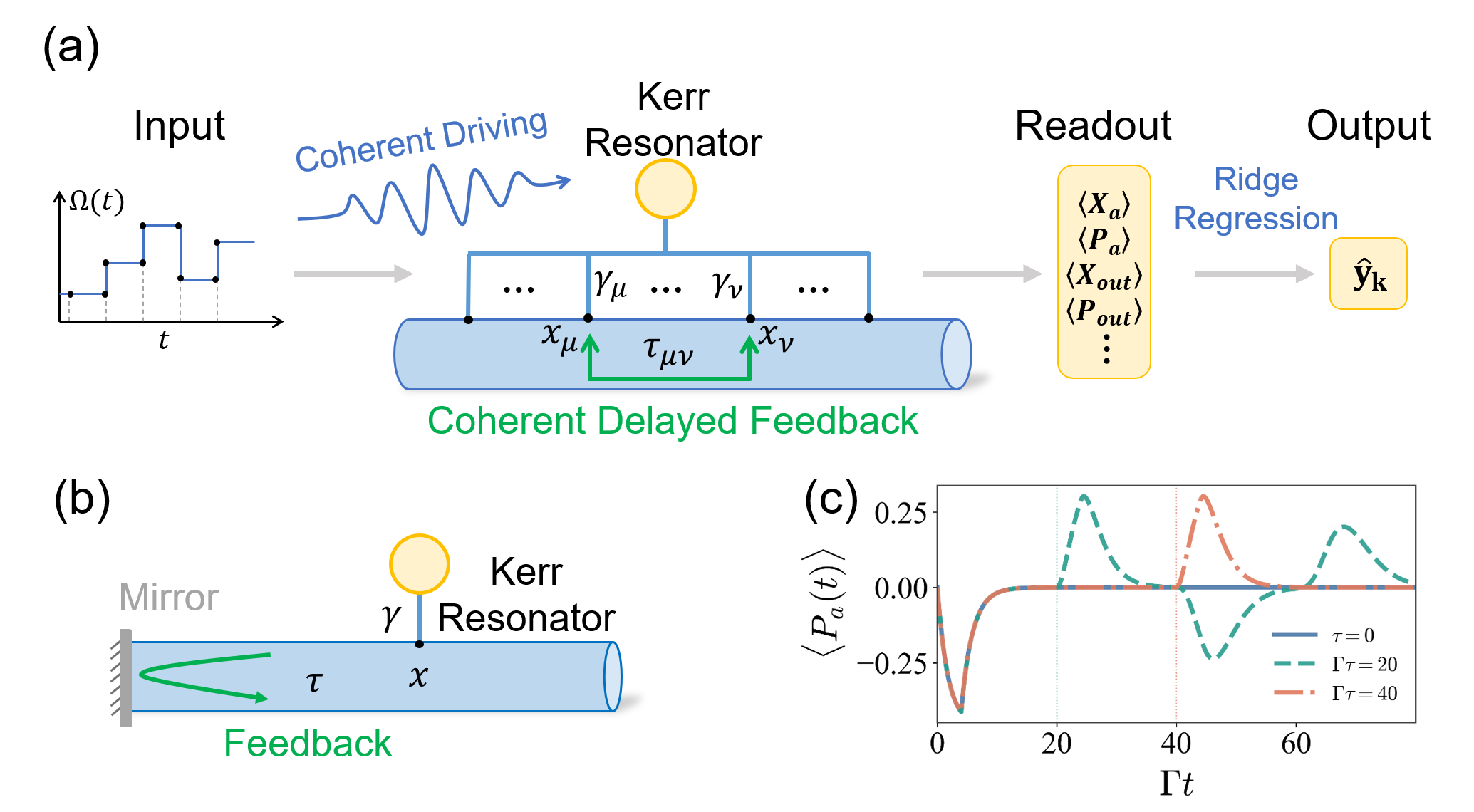}
    \caption{
    Non-Markovian waveguide-QED reservoir.
    (a) A driven Kerr-resonator couples to a 1D waveguide at multiple points, generating coherent delayed feedback with propagation times $\tau_{\mu\nu}$. The drive $\Omega(t)$ encodes the input, while the measured reservoir features are mapped to $\hat y_k$ through a trained linear readout.
    (b) Minimal single-delay geometry  equivalently represented by a resonator coupled to a semi-infinite waveguide, with coherent memory-return delay $\tau$.
    (c) Resonator dynamics
    following a short-pulse excitation for representative delays $\Gamma \tau=\{0,20,40\}$, illustrating the delay-controlled return of information to the resonator.
    }
    \label{fig:model}
\end{figure}

Measurement-feedback QRC schemes offer an explicit route to memory reinjection, but typically rely on repeated measurements and the classical processing of the outcomes~\cite{kobayashi2024,zhu2025,dibartolo2026,paparelle2026}. 
Non-Markovian quantum reservoirs, by contrast, can preserve coherent dynamics
and enhance memory without intermediate measurement, yet their temporal memory profiles are typically fixed by the intrinsic reservoir dynamics rather than engineered for specific task-relevant timescales~\cite{sannia2026,sasaki2025}.
The central challenge is therefore to engineer the temporal
structure of quantum-reservoir memory through coherent feedback,
without intermediate measurement or classical reinjection, and
to match the resulting memory profile to task-relevant temporal
dependencies.

In this work, we propose
a non-Markovian waveguide-QED reservoir consisting of a driven giant Kerr-nonlinear resonator coupled nonlocally to a one-dimensional (1D) waveguide at spatially seperated
points, see Fig. \ref{fig:model}.
The required ingredients
---nonlocal coupling, finite propagation times, and nonlinear resonator dynamics---are compatible with existing giant-atom and waveguide-QED platforms~\cite{kockum2018,andersson2019,kannan2020,vadiraj2021,xiao2025}.
The emitted field carries information away from the resonator and coherently
returns it after a geometry-defined delay, so that memory is encoded in the
joint resonator--waveguide dynamics rather than stored solely in the local
resonator.
This architecture realizes a quantum time-delay reservoir in which the unmeasured propagating field serves as a coherent memory carrier, returning information about past inputs to the nonlinear resonator.
Waveguide propagation determines when information about past inputs becomes dynamically available again, whereas the Kerr interaction mixes the returning field with the current resonator state and converts it into nonlinear temporal features accessible through a linear readout. Delay-resolved STM  and delayed-product capacities identify these two roles independently. We further show that, in NARMA-$n$ prediction, the optimal feedback delay shifts systematically with task order and tracks the explicit input-product lag. These results establish waveguide-mediated coherent feedback as a hardware-level mechanism for engineering task-relevant memory in QRC.

The remainder of this paper is organized as follows. In Sec.~\ref{sec:model}, we introduce the waveguide-QED reservoir architecture, derive the effective geometry-induced memory kernel, and present the input encoding and readout protocol. Sec.~\ref{sec:memory} uses delay-resolved benchmark tasks to disentangle the complementary roles of coherent propagation and Kerr nonlinearity in linear and nonlinear memory formation. In Sec.~\ref{sec:prediction}, we demonstrate that this programmable memory profile enables task-matched nonlinear prediction using NARMA-$n$ benchmarks. Finally, Sec.~\ref{sec:discussion} summarizes the main conclusions and discusses future directions for geometry-programmed QRC.

\section{Waveguide-QED reservoir and Learning Protocols}
\label{sec:model}

\subsection{Hamiltonian model and memory kernel}

We consider the waveguide-QED reservoir illustrated in Fig.~\ref{fig:model}(a), consisting of a driven
Kerr-nonlinear resonator coupled to a 1D waveguide at $M$
spatially separated coupling points $x_1,\ldots,x_M$. 
In a frame rotating at the drive frequency $\omega_d$, and in the interaction picture with respect to the free waveguide Hamiltonian, the total Hamiltonian reads
\begin{align}
H(t)
&=
\Delta a^\dagger a
+\frac{U}{2}a^{\dagger 2}a^2
+i\!\left[\Omega(t)a^\dagger-\Omega^*(t)a\right]
\notag\\
&\quad+
\sum_{\mu=1}^{M}\sum_{\alpha=L,R}
\int d\omega\,
\left[
g_{\mu\alpha\omega}e^{-i\delta_\omega t}
a^\dagger b_\alpha(\omega)
+{\rm H.c.}
\right].
\label{eq:full_hamiltonian}
\end{align}
where $a$ annihilates an excitation of the resonator and $b_\alpha(\omega)$ annihilates a waveguide mode of frequency $\omega$ propagating in direction $\alpha=L,R$. The parameters $\Delta=\omega_r-\omega_d$ and $\delta_\omega=\omega-\omega_d$ denote the resonator and waveguide detunings in the rotating frame, respectively, $U$ is the Kerr nonlinearity, and $\Omega(t)$ is the input-dependent coherent drive. For symmetric coupling to left- and right-propagating modes, we take
$g_{\mu\alpha\omega}=\sqrt{\gamma_\mu/4\pi}\,
e^{is_\alpha\omega x_\mu/v_g}$, with $s_R=+1$ and $s_L=-1$, so that $\gamma_\mu$ is the total radiative decay rate of coupling point $\mu$.

Eliminating
the waveguide degrees of freedom yields a nonlocal 
Heisenberg-Langevin equation for the resonator operator~\cite{barahona-pascual2026}. For resonant driving, ($\Delta=0$), it takes the form
\begin{equation}
    \dot a(t)
    =
    -iU a^\dagger(t)a^2(t)
    -
    \int_0^t K(t-t')a(t')\,dt'
    +
    \Omega(t)
    +
    \xi(t),
    \label{eq:delayed_dynamics}
\end{equation}
where $\xi(t)$ is the zero-mean vacuum input noise and $K(t)$ is the (effective) memory
kernel generated by waveguide propagation. Within a local Markov approximation at each coupling point, while retaining the
finite propagation delays between spatially separated points~\cite{grimsmo2015,barahona-pascual2026,chen2026},
the  memory kernel can be written as
\begin{equation}
    K(t)
    \simeq
    \frac{\Gamma}{2}\delta(t)
    +
    \sum_{\mu\neq\nu}
    \frac{\sqrt{\gamma_\mu\gamma_\nu}}{2}
    e^{i\phi_{\mu\nu}}
    \delta(t-\tau_{\mu\nu}) ,
    \label{eq:geometry_kernel}
\end{equation}
where $\Gamma=\sum_\mu\gamma_\mu$ is the total radiative decay rate,
$\tau_{\mu\nu}=|x_\mu-x_\nu|/v_g$ is the propagation time between coupling
points, and $\phi_{\mu\nu}\simeq\omega_d\tau_{\mu\nu}$ is the corresponding
propagation phase. The first, instantaneous  term in Eq.~\eqref{eq:geometry_kernel} describes the local Markovian decay, whereas each delayed term represents a geometry-induced coherent feedback channel. 
Through this channel, a field emitted by the resonator propagates through the waveguide and coherently re-enters
the resonator after a finite propagation time
\(\tau_{\mu\nu}\), with feedback strength
\(\sqrt{\gamma_\mu\gamma_\nu}/2\), without intermediate measurement
or classical reinjection..

To isolate the elementary mechanism of waveguide-mediated coherent memory, we focus on the minimal nontrivial configuration consisting of two
coupling points separated by a single propagation delay $\tau=\tau_{12}$.
Despite its simplicity, this geometry already captures the complete
physical cycle underlying the proposed reservoir: information carried by an unmeasured propagating field leaves the
resonator, is temporarily stored in the waveguide, returns after a
finite propagation time, and is subsequently transformed by the
local Kerr nonlinearity into nonlinear temporal features.
It therefore provides the simplest setting in which memory transport
can be cleanly separated from memory processing. The same effective single-delay feedback channel can be implemented  experimentally by placing
the resonator in front of a mirror, or equivalently by coupling it to a
semi-infinite waveguide, as illustrated in Fig.~\ref{fig:model}(b). In this representation, the feedback delay is simply determined by
the resonator-mirror distance.
Following a short-pulse excitation, the returning field produces a
delayed echo in the resonator response
[Fig.~\ref{fig:model}(c)], providing a direct dynamical signature of
the coherent memory channel.
The multi-point geometry considered in Eq.~\eqref{eq:geometry_kernel}
extends this minimal construction to multiple independently
programmable memory-return times.

The effective kernel in Eq.~\eqref{eq:geometry_kernel} naturally separates
two physical ingredients of the reservoir. Waveguide propagation determines
when information carried by the emitted field coherently returns to the
resonator, whereas the Kerr interaction determines how this returning field is
transformed into nonlinear computational features. Relative to the input
update interval \(\Delta t\), the ratio \(\tau/\Delta t\) fixes the temporal
location of this coherent memory return in discrete input steps. To distinguish these two roles, we compare a time-delay linear reference
reservoir (TDLRC, $U=0$), which isolates coherent memory transport, with a
time-delay Kerr quantum reservoir (TDQRC, $U=30\,\Gamma$), in which coherent
memory transport and nonlinear memory processing coexist. All remaining
parameters are kept identical. 
The TDLRC is simulated by solving the corresponding delayed-differential equation (DDEs)~\cite{barahona-pascual2026}, while the TDQRC is simulated using the time-bin matrix-product-state (MPS) method~\cite{pichler2016,regidor2026,garcia-molina2026}, which explicitly respresents the prorogating waveguide field responsible for
coherent delayed feedback. Implementation details are provided in Appendix~\ref{app:mps}. Unless stated
otherwise, all results are obtained for balanced couplings $\gamma_1=\gamma_2=\Gamma/2$
and constructive propagation phases $\phi_{\mu\nu}=2\pi m$. The effects of propagation phase, coupling asymmetry,
non-integer delays, and multi-delay geometries are discussed in the
Appendices~\ref{app:phase_scan}
and~\ref{app:gamma_scan}.

Together, the geometry-induced memory kernel, coherent input encoding, and
linear readout establish a reservoir in which coherent waveguide propagation
controls the temporal placement of memory, while intrinsic Kerr dynamics
govern its nonlinear processing. The following sections quantify these two
roles independently using delay-resolved memory benchmarks.

\subsection{Input encoding and readout}

The reservoir is driven by a scalar input sequence $\{u_k\}$, encoded in the coherent drive by zero-order-hold protocol, \begin{equation}
    \Omega(t)=\epsilon u_k,
    \qquad
    (k-1)\Delta t\le t<k\Delta t ,
    \label{eq:input_encoding}
\end{equation}
where $\epsilon$ sets the input amplitude and $\Delta t$ is the input update interval  Each input value is therefore held constant for one
reservoir-evolution step, while the waveguide-mediated dynamics retain and
process information from preceding steps.

The reservoir state is accessed through first-order quadrature moments of
both the local resonator and the outgoing waveguide field,
\begin{equation}
    \mathbf r(t)
    =
    \big(
    \langle X_a(t)\rangle,\,
    \langle P_a(t)\rangle,\,
    \langle X_{\rm out}(t)\rangle,\,
    \langle P_{\rm out}(t)\rangle
    \big)^{\mathbf{T}} ,
    \label{eq:response_vector}
\end{equation}
where $X_a=a+a^\dagger$, $P_a=i(a^\dagger-a)$, $X_{out}=b_{out}+b_{out}^\dagger$ and $P_{out}=i(b_{out}^\dagger-b_{out})$, and $b_{out}$ is the waveguide output field operator.  Because the readout
contains only first-order moments and is subsequently processed by a linear
regression layer, any nonlinear temporal features accessible to the output
must be generated by the intrinsic reservoir dynamics. 

In order to enrich the accessible dynamical features without introducing an additional memory channel, we adopt temporal multiplexing~\cite{Appeltant2011}. During each input interval $k$, the response is sampled at $t_{k,m}=k\Delta t-m\theta$ ($m=0,\dots,N_{\rm v}-1$), where $\theta$ is the virtual-node spacing and $N_{\rm v}$ is the number
of virtual nodes. The resulting feature vector is 
\begin{equation}
    \mathbf x_k
    =
    \left(
    1,\,
    \mathbf r^{\mathbf{T}}(t_{k,0}),\dots,
    \mathbf r^{\mathbf{T}}(t_{k,N_{\rm v}-1})
    \right)^{\mathbf{T}},
    \label{eq:feature_vector}
\end{equation}
where the leading constant provides the readout bias.
We impose $(N_{\rm v}-1)\theta\le\Delta t$ 
so that all virtual nodes lie within a single input interval. Temporal
multiplexing therefore resolves the intra-step reservoir trajectory but does
not explicitly embed responses from earlier input intervals. Memory extending
beyond one input step must consequently originate from the resonator--waveguide
dynamics. 
The predicted output is obtained from the linear readout
\begin{equation}
    \hat y_k=W_{\rm out}\mathbf x_k .
    \label{eq:linear_readout}
\end{equation}
The weights $W_{\rm out}$ are trained (and thus optimized) via ridge regression~\cite{vrugt2026} after discarding an initial washout period, and are evaluated on an independent
test data. The resonator is initially
in its ground state and the waveguide is in vacuum. The decay of feature-level
dependence on the initial resonator state, and hence the echo-state property in
the operating regime considered here, is verified in
Appendix~\ref{app:esp}.

\section{Linear and Nonlinear Memory}
\label{sec:memory}

Having established the geometry-induced memory kernel in Sec.~\ref{sec:model}, 
we now use delay-resolved benchmark tasks to identify the distinct roles of coherent waveguide propagation and Kerr nonlinearity. The short-term-memory task resolves when information about a past input
becomes accessible to the readout, whereas the delayed-product task tests
whether this returned information is nonlinearly combined with the current
input. 

The STM target is
\begin{equation}
    y_{k,{\rm STM}}^{(d)} = u_{k-d},
\end{equation}
where $d$ is the input lag and the random inputs
are independently sampled from $u_k\sim\mathcal{U}(0,1)$. To probe nonlinear delayed-memory processing, we
use the delayed-product target
\begin{equation}
    y_{k,{\rm prod}}^{(d)}
    =
    (u_k-\bar u)(u_{k-d}-\bar u),
\end{equation}
where $\bar u$ is the input mean. This task requires both the retention of $u_{k-d}$ 
and its multiplicative mixing with the current input $u_k$. For each task,
the delay-resolved capacity is defined as
\begin{equation}
    C_d^{\eta}
    =
    \frac{
    {\rm Cov}^2
    (\hat y_{k,\eta}^{(d)},y_{k,\eta}^{(d)})
    }{
    {\rm Var}(\hat y_{k,\eta}^{(d)})
    {\rm Var}(y_{k,\eta}^{(d)})
    },
    \quad
    \eta\in\{{\rm STM},{\rm prod}\},
    \label{eq:capacity_def}
\end{equation}
with all quantities evaluated on held-out test data after washout and training. The capacity satisfies $0\le C_d^\eta\le 1$, with unity corresponding to perfect linear reconstruction of the target from the
reservoir features.

\begin{figure}[tp]
    \centering
    \includegraphics[width=\linewidth]{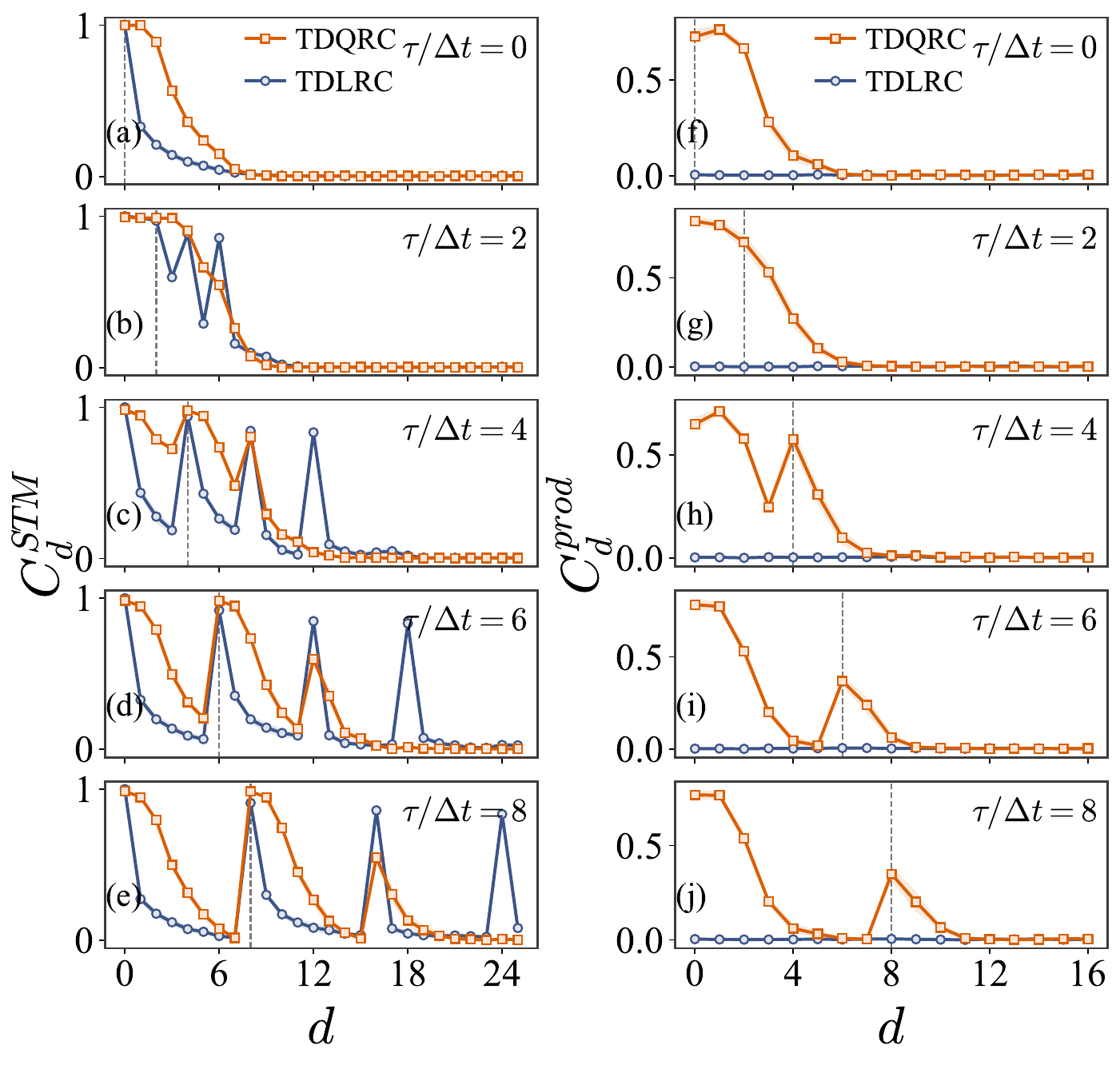}
    \caption{
    Delay-resolved linear and nonlinear memory. The left and right columns show
    the STM capacity $C_d^{\rm STM}$ for reconstructing $u_{k-d}$ and the
    delayed-product capacity $C_d^{\rm prod}$ for reconstructing
    $(u_k-\bar u)(u_{k-d}-\bar u)$, respectively. Rows correspond to feedback
    delays $\tau/\Delta t=0,2,4,6,8$;
    vertical dashed lines mark
    $d=\tau/\Delta t$. TDQRC denotes the Kerr-nonlinear time-delay quantum reservoir, while TDLRC the linear reference reservior with the
    same feedback geometry. 
    Parameters are $\Delta t=2$, $\epsilon=1$,
    $\gamma_1=\gamma_2=0.1$ ($\Gamma=0.2$), $N_{\rm v}=20$, and $\theta=0.1$.
    Washout, training, and test sequences contain 100, 600, 200 input steps, respectively. Shaded regions indicate one standard deviation over independent input realizations.
    }
    \label{fig:memory_product}
\end{figure}

The STM capacity directly resolves the  temporal memory structure induced
by the
memory kernel in Eq. \eqref{eq:geometry_kernel}. 
In the absence of a finite
feedback delay, the instantaneous kernel contribution produces a local
fading-memory envelope, as seen for $\tau/\Delta t=0$ in Fig.~\ref{fig:memory_product}(a). 
In this regime, the TDQRC exhibits a
broader short-lag response
than the TDLRC, indicating that the Kerr dynamics
reshape the local reservoir relaxation even before a temporally separated return is
introduced.

For finite $\tau$, information emitted into waveguide coherently re-enters the resonator after the propagation time. Relative to the input
update interval $\delta(t-\tau)$, this produces a primary memory revival near
\begin{equation}
    d_{\rm \tau}\simeq \frac{\tau}{\Delta t}.
    \label{eq:primary_revival_law}
\end{equation}
At the shortest finite delay
[Fig.~\ref{fig:memory_product}(b)], the
returning field overlaps with the
local fading-memory response, resulting in
a broadens low-lag profile.
As the delay increases
[Fig.~\ref{fig:memory_product}(c)-(e)], the
returned component separates from the local response and forms a distinct STM revival.
Its position follows Eq.~\eqref{eq:primary_revival_law} in both the TDLRC and
TDQRC, demonstrating that the timing of the primary memory return is governed
mainly by waveguide propagation rather than by the local nonlinearity.  

In addition, secondary revivals appear
at later lags as a consequence of repeated
coherent reinteractions between the resonator and the delayed field. In the
parameter regime considered here, these echoes are more pronounced
in the TDLRC, where the returning field evolves linearly and can accumulate coherently over successive round trips.
In the TDQRC, Kerr nonlinearity mixes such returning components
with the instantaneous resonator state, thereby reshaping the echo profiles and reducing the amount of
information that remains accessible as purely linear delayed recall.
This
behavior reflects the familiar redistribution of reservoir capacity between
linear memory and nonlinear processing
~\cite{dambre2012a,verstraeten2010,xia2023a,cindrak2026a}. Waveguide
propagation therefore sets the temporal positions of the coherent returns,
whereas Kerr dynamics redistribute their information content among linear and
nonlinear features. 

Next, we use the delayed-product task to determine whether  the coherently
returned information is converted into nonlinear temporal features. 
Despite its clear STM revivals, the TDLRC exhibits negligible $C_d^{\rm prod}$ under the present first-order-moment and linear-readout protocol [Fig.~\ref{fig:memory_product}(f)--(j)]. A linear reservoir can retain $u_{k-d}$, but it cannot efficiently generate the multiplicative
feature $u_k u_{k-d}$ from the same first-order observables.
The TDQRC behaves qualitatively differently.
At the short
finite delay [Fig.~\ref{fig:memory_product}(g)], the delayed-product
response extends over
over neighboring lags, because  the coherent return overlaps with the local
nonlinear response.
For larger delays,
the two contributions become
temporally separated, and a distinct product-capacity feature emerges near
$d\simeq\tau/\Delta t$, see Eq.~\eqref{eq:primary_revival_law}.
The coincidence between the STM and
delayed-product revival positions shows  that the waveguide first transports
information about \(u_{k-d}\) back to the resonator, after which the Kerr
interaction combines it with the current reservoir response and renders the
resulting nonlinear feature accessible to a linear readout.
The integer values of $\tau/\Delta t$ used in Fig.~\ref{fig:memory_product} provide the transparent illustration of delay-programmed memory.  More generally, when
$\tau/\Delta t$ is non-integer, the coherent memory return is distributed
across the nearest discrete input lags rather than being localized at a
single delay, as discussed in
Appendix~\ref{app:noninteger-delay}. Likewise, extending the giant resonator
to multiple coupling points introduces several coherent feedback paths and
thus multiple programmable memory-return times, whose resulting memory
structure is analyzed in Appendix~\ref{app:multi_delay}.

These benchmarks establish a two-stage mechanism for delay-based quantum reservoir processing. Coherent waveguide propagation
controls when past-input information becomes available again, while the Kerr
interaction determines whether that information remains accessible as linear
recall or is transformed into nonlinear delayed features. This separation
suggests a direct task-matching principle: temporal prediction should improve
when a coherent memory return is positioned near a task-relevant lag and is
processed by sufficiently strong intrinsic nonlinearity. We test this
principle next using NARMA-$n$ prediction.

\section{Task-Matched Time-Series Prediction}
\label{sec:prediction}

Now, we test whether the delay-resolved memory mechanism identified in
Sec.~\ref{sec:memory} can be translated into improved nonlinear time-series
prediction. In contrast to the STM and delayed-product benchmarks, which probe
specific temporal features at a prescribed lag, nonlinear forecasting tasks
combine multiple delayed dependencies and recurrent nonlinear transformations.
They therefore provide a more stringent test of whether coherent memory return
can be matched to task-relevant temporal structure.

Prediction accuracy
is quantified by the normalized root-mean-square error (NRMSE):
\begin{equation}
    \mathrm{NRMSE}
    =
    \left[
    \frac{\sum_k (y_k-\hat y_k)^2}
    {\sum_k (y_k-\bar y)^2}
    \right]^{1/2},
    \label{eq:nrmse}
\end{equation}
where $\bar y$ is the mean value of target outputs over the test set.

We consider the NARMA-$n$ benchmark, which is widely used to assess the
combined nonlinear-processing and finite-memory capabilities of reservoir
computers~\cite{Appeltant2011}.
The target sequence obeys
\begin{equation}
    y_{k+1}
    =
    0.3\,y_k
    +
    0.05\,y_k
    \sum_{j=0}^{n-1} y_{k-j}
    +
    1.5\,\tilde u_k\,\tilde u_{k-n+1}
    +
    0.1 ,
    \label{eq:narma}
\end{equation}
with $\tilde u_k=u_k/5\in[0,0.2]$. The explicit product
$\tilde u_k\tilde u_{k-n+1}$
couples the current input to
a past input at the lag $d_{\rm task}=n-1$. 
The recurrent terms involving the previous outputs $y_k$ 
introduce additional, distributed temporal dependencies, but the product term provides
a clearly indentifiable
nonlinear input-input correlation that 
can be directly compared
with the delay-induced nonlinear memory characterized in
Sec.~\ref{sec:memory}.

According to Eq.~\eqref{eq:primary_revival_law}, coherent delayed feedback returns 
past-input information to the resonator near
$d_\tau \simeq  \tau/\Delta t$.
The Kerr nonlinearity can then mix this returned information with the response to current input,
producing nonlinear, Volterra-type
temporal features of the form $u_k u_{k-d_\tau}$. Matching this feature to the explicit NARMA product thus suggest an optimal task-matching
condition
\begin{equation}
    \frac{\tau_{\rm opt}}{\Delta t}
    \simeq
    d_{\rm task}
    =
    n-1 .
    \label{eq:narma_matching_law}
\end{equation}
Because NARMA-$n$ also contains recurrent and distributed memory terms,
Eq.~\eqref{eq:narma_matching_law} should be understood as a physically
motivated design rule rather than an exact identity.

\begin{figure}[t]
\centering
\includegraphics[width=\linewidth]{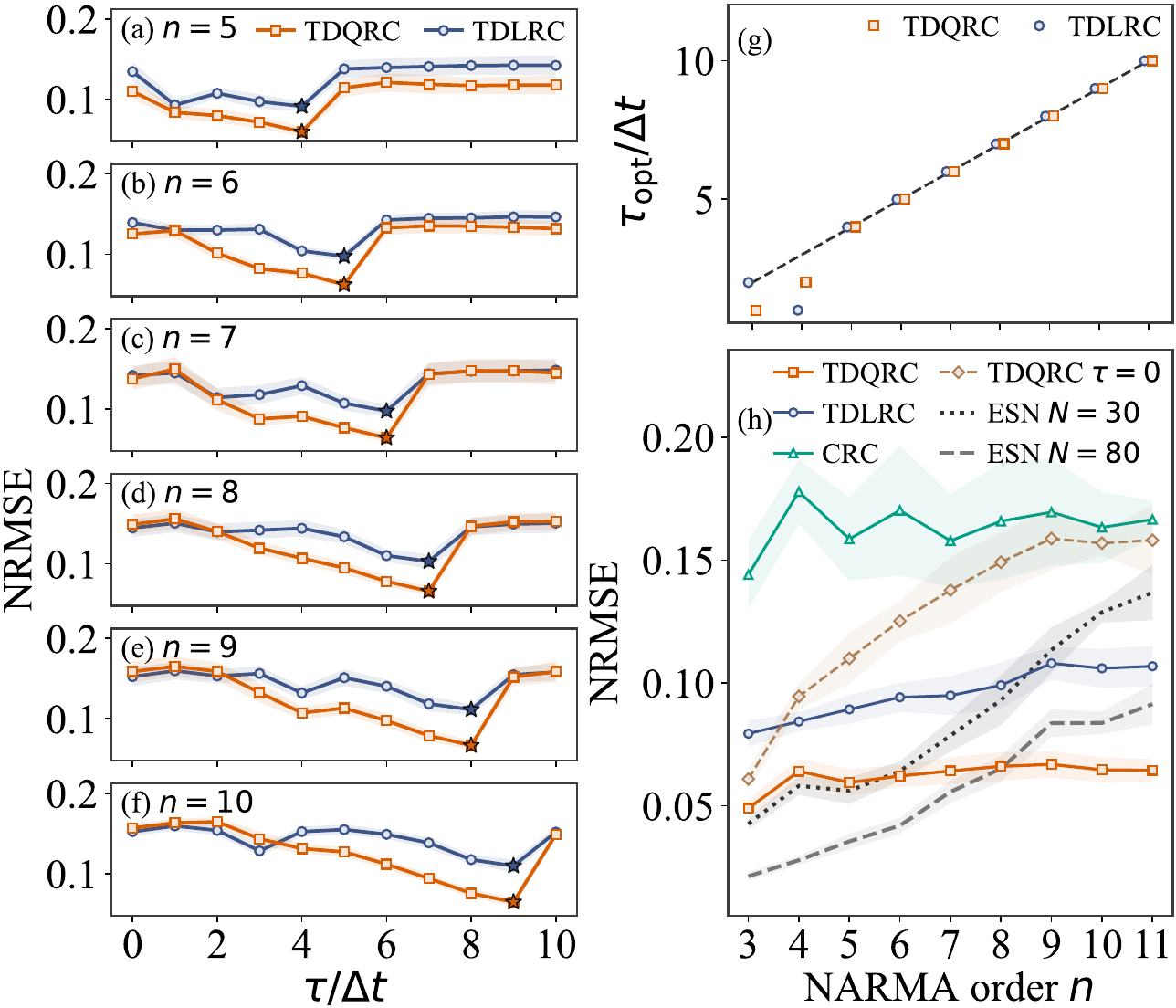}
\caption{
Task-matched prediction results on NARMA-$n$ sequences.
(a)-(f) Test NRMSE as a function of programmed delay \(\tau/\Delta t\) for  \(n\in\{5,6,7,8,9,10\}\), comparing TDQRC and TDLRC. Stars indicate the minimum NRMSE for each curve. (g) Extracted optimal delay \(\tau_{\rm opt}/\Delta t\) versus the explicit input-product lag \(n-1\). The dashed identity line represents the matching condition in Eq.~\eqref{eq:narma_matching_law}. (h) NRMSE as a function of NARMA order for the reservoir designs and baselines considered. All parameters are the same as in
Fig.~\ref{fig:memory_product}.
}
\label{fig:narma}
\end{figure}

As the NARMA order increases, the delay yielding the minimum prediction error shifts systematically toward larger values [Fig.~\ref{fig:narma}(a)-(f)]. The extracted optimal delays closely track the explicit produce lag \(n-1\) across the tested orders [Fig.~\ref{fig:narma}(g)], supporting the condition in Eq.~\eqref{eq:narma_matching_law}.
This trend shows that the predication improves when the coherent
memory return is positoned near  near a temporal correlation
explicitly required by the target dynamics.
Delays that are too short return
information before the relevant past input is required, whereas excessively
long delays shift the strongest memory revival beyond this task-relevant lag.
This interpretation is independently supported by the delayed-product
capacity evaluated at \(d=n-1\). Its delay dependence selects the same
feedback region as the NARMA error minimum, as shown in
Appendix~\ref{app:narma_product_matching}. The agreement connects the
task-level prediction optimum directly to the nonlinear memory feature
identified in Sec.~\ref{sec:memory}.

Control comparisons
separate the
roles of coherent memory placement and nonlinear processing.
The TDLRC retains a clear delay-dependent error
landscape, demonstrating that the temporal placement of memory alone already
affects prediction. However, because its first-order observables remain linear
functions of the input history, it cannot efficiently generate the nonlinear
delayed features required by the NARMA product term. At matched delay, the
TDQRC therefore achieves lower errors, with the advantage becoming more
pronounced as the NARMA order increases.
The zero-delay QRC represents the complementary limit: it retains Kerr nonlinearity but lacks a temporally separated coherent memory return. Its error increases
rapidly with NARMA order [Fig.~\ref{fig:narma}(h)], showing that
local nonlinear processing alone is insufficient when the task requires
longer-range temporal correlations. High-order NARMA prediction thus benifits from the coexistence of geometry-controlled memory return and Kerr-induced
nonlinear mixing.

Additional baseline test whether the observed improvement can be attributed
solely to generic classical nonlinearity or reservoir size.
Under the same delayed-feedback kernel readout protocol, a scalar classical nonlinear resonator (CRC) does not reproduce the full nonlinear delayed-memory response of the TDQRC; detailed comparisons are provided in Appendix~\ref{app:crc}. Standard echo-state networks (ESNs)~\cite{zhu2025,zhu2025b,hou2026} remains competitive at low NARMA orders but their error increase more rapidly with $n$ in the parameter and resouce regime considered here. Among the tested configurations, the
matched-delay TDQRC gives the best high-order NARMA performance
[Fig.~\ref{fig:narma}(h)].

These results establish that NARMA predication performance is governed not by
memory depth alone, but by the alignment between the temporal structure of the
reservoir memory and that of the target task. For NARMA-\(n\), the explicit
input-product lag provides a transparent design target, and the optimal
feedback delay follows this lag closely.
The boarder Mackey--Glass forecasting task, analyzed in Appendix~\ref{app:mg}, contains distributed rather than single-lag temporal correlations.  Its delay dependence is correspondingly less sharply localized, but it likewise benefits from the combination of
coherent delayed feedback and Kerr nonlinear processing.

\section{Discussion and Conclusion}
\label{sec:discussion}

We have demonstrated that coherent delayed feedback in waveguide QED provides a direct physical mechanism for engineering the temporal memory structure of a
quantum reservoir.  Information about past inputs is carried away from the
nonlinear resonator by an unmeasured propagating field and coherently returned
after a geometry-controlled propagation time. The waveguide therefore
determines when past information becomes available again, while the local Kerr
interaction determines how this returned information is transformed into
nonlinear temporal features accessible through a linear readout. Unlike
measurement-based feedback schemes, this memory channel requires neither
intermediate detection nor classical reinjection.

The delay-resolved STM and delayed-product benchmarks directly reveal these
complementary roles. The primary linear-memory revival appears near \(d_\tau \simeq \tau/\Delta t\), showing that the temporal position of the returned information is controlled by the propagation delay. The corresponding delayed-product response shows
that Kerr dynamics convert this coherently returned information into
nonlinear features involving both past and current inputs. These observations
lead to a task-matching principle: reservoir performance depends not only on
the total amount or depth of memory, but also on whether the memory is
available at lags relevant to the target dynamics. The NARMA-\(n\) results provide a direct demonstration of this principle. The optimal feedback delay closely follows
the optimal condition \eqref{eq:narma_matching_law}. The agreement between the task lag, the delayed-product-capacity maximum, and the minimum prediction error connects the computational optimum to a specific physical feature of the reservoir memory kernel. The resulting improvement is therefore not attributable to longer memory alone: it arises from aligning a coherent memory return with a task-relevant nonlinear temporal dependence.

The present study focuses on a minimal single-delay geometry, first-order
quadrature readout, and a local Kerr nonlinearity. These choices make it
possible to isolate the elementary mechanisms of coherent memory transport and
nonlinear memory processing, but they do not exhaust the design space of
waveguide-QED reservoirs. Multipoint coupling can generate several
independently structured return times, while propagation phases and coupling
asymmetries provide additional control over the amplitudes and interference of
the corresponding memory channels. Loss, dephasing, and imperfect feedback
will determine how much of this coherent memory remains accessible in realistic
devices. More generally, networks of coupled non-Markovian nodes may support
distributed and hierarchical memory kernels that cannot be realized with a
single resonator.

Last but not least, waveguide-QED and giant-atom platforms are particularly suited to this
approach because spatially separated coupling points, finite propagation
times, interference phases, and strong local nonlinearities can be engineered
within the same architecture. More broadly, superconducting circuit-QED
platforms \cite{bertet2012circuit,kannan2020waveguide,PhysRevA.90.013837,l1fq-gbbl} provide a natural implementation route, where Josephson
Kerr resonators, giant artificial atoms, and microwave waveguides can be
integrated with high tunability and high-fidelity quadrature readout. 
Our results therefore establish geometry-programmed coherent memory as a hardware-level design principle for non-Markovian quantum information
processing, with QRC providing one concrete
application.

\section*{Data availability}
Data and code are available from the authors upon reasonable request.

\begin{acknowledgments}
  This work was supported by CSIC's JAE Chair Program 2024 and PRO-ERC AGAIN 2025 funding. H.J. gratefully acknowledges financial support from the China Scholarship Council (CSC202308620117). Y.B. acknowledges support from the Spanish Ministry of Science, Innovation and Universities through Grant PID2024-157842OA-I00 and from the Spanish national project in Artificial Intelligence through Grant AIA2025-163435-C44. X.C. acknowledges support from the Spanish Ministry of Science, Innovation and Universities through Grant PID2021-126273NB-I00. The authors also acknowledge support from the Severo Ochoa Centres of Excellence programme through Grant CEX2024-001445-S. 
  
\end{acknowledgments}

\appendix

\section{Numerical methods and system-field correlations}
\label{app:mps}

\begin{figure}[t]
    \centering
    \includegraphics[width=\linewidth]{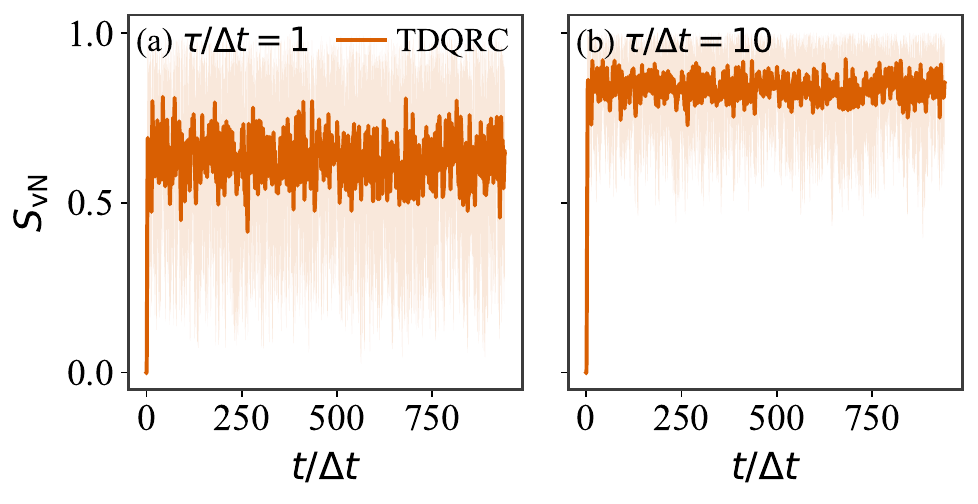}
    \caption{
    Schmidt entropy during the STM task for representative feedback delays. The entropy is evaluated across the bipartition between the resonator and the active delay-line modes in the time-bin MPS.
    }
    \label{fig:app_entropy}
\end{figure}
To separate coherent memory transport from nonlinear memory processing, we
employ two complementary numerical descriptions for the linear and nonlinear
reservoirs introduced in the main text.
For the time-delay linear reference reservoir (TDLRC), we replace the
resonator operator in Eq.~\eqref{eq:delayed_dynamics} by a complex field
amplitude $a(t)\rightarrow \alpha(t)$ and set the Kerr nonlinearity to zero. For the two-point geometry with constructive feedback, the resulting delay-differential equation is
\begin{align}
    \dot\alpha(t)
    =
    -\frac{\Gamma}{2}\alpha(t)
    -
    \frac{\sqrt{\gamma_1\gamma_2}}{2}\alpha(t-\tau)
    +
    \Omega(t),
    \label{eq:crc_two_point}
\end{align}
with the same memory-kernel convention as in Eq.~\eqref{eq:geometry_kernel}. This description retains the effect of
waveguide propagation through the delayed amplitude \(\alpha(t-\tau)\), while
eliminating the waveguide degrees of freedom explicitly.

For the time-delay Kerr quantum reservoir (TDQRC), we simulate the joint
resonator-waveguide quantum dynamics using the time-bin matrix-product-state
(MPS) method of Ref.~\cite{pichler2016}, implemented with the MPS libraries of
Refs.~\cite{regidor2026,garcia-molina2026}. Unlike the delay-differential
description, the time-bin MPS explicitly retains the propagating waveguide
degrees of freedom that carry the coherent non-Markovian memory. The waveguide field is discretized into temporal modes of duration
\(\Delta t_{\rm bin}\). A propagation delay is represented by an integer
number of time bins,
$\tau_{\mu\nu}=\ell_{\mu\nu}\Delta t_{\rm bin}$, so that a field bin emitted at one coupling point reinteracts with the
resonator after \(\ell_{\mu\nu}\) time steps.  

In the time-bin representation, the waveguide is mapped onto an ordered chain
of field modes. At
each time step, the resonator interacts with the bin, or bins, arriving at its
coupling points. The relevant field modes are
then advanced along the MPS chain, and the state is recompressed by singular
value decomposition. This construction explicitly retains the temporally
separated field modes that mediate coherent delayed feedback. Reservoir
features are obtained from resonator observables and from outgoing time-bin
observables after their final interaction with the resonator.

For the symmetric constructive-feedback geometries considered in the main
text,  we
use an equivalent folded chiral representation to reduce the numerical cost. In this construction, the left- and right-propagating
waveguide modes are mapped onto a single effective chiral channel while
preserving the coherent delayed interaction.
For the two-point geometry, the
resulting model is equivalent to a unidirectional delay line,
or to a Kerr resonator coupled to a semi-infinite waveguide in front of a mirror~\cite{pichler2016}. 

Unless stated
otherwise, the TDQRC simulations use a time-bin duration $\Delta t_{\rm bin}=0.1$,  a local  time-bin cutoff
$d_{\rm bin}=4$, a maximum MPS bond dimension \(\chi=24\), and a resonator Hilbert-space cutoff
\(d_{\rm sys}=12\). Numerical convergence was verified with respect to
\(\Delta t_{\rm bin}\), \(d_{\rm bin}\), \(\chi\), and \(d_{\rm sys}\). The
reported truncation parameters were chosen such that further increases produce
no appreciable changes in the reservoir observables or benchmark performance.

To verify that the delayed memory is carried by coherent system-field correlations rather than by an effective classical delay variable,, we monitor the Schmidt entropy during the STM task. From the time-bin MPS, we extract the Schmidt
coefficients $\{\lambda_\alpha\}$ across the bipartition between the resonator and the active delay-line modes
and compute
\begin{align}
    S_{\rm vN}
    =
    -\sum_\alpha
    \lambda_\alpha^2
    \log_2 \lambda_\alpha^2 ,
    \label{eq:schmidt_entropy}
\end{align}
where the normalized Schmidt weights satisfy
\(\sum_\alpha\lambda_\alpha^2=1\).
Fig.~\ref{fig:app_entropy} shows that the TDQRC develops a finite Schmidt entropy, with representative values $S_{\rm vN}\sim0.6$--$0.9$ bits.
The non-zero
Schmidt entropy demonstrates that the resonator becomes entangled with the  waveguide modes retained inside the active delay line. The memory channel
studied in the main text is therefore physically encoded in the joint system--field state carried by the propagating waveguide modes, rather than in
a classical delayed variable.

\section{Propagation-phase dependence}
Here we examine how the phase accumulated by the field
during propagation modifies the delay-resolved memory response. For the two-point geometry, the memory kernel may be written schematically as
\begin{equation}
    K(t)
    =
    K_0\delta(t)
    +
    K_\tau e^{i\phi}\delta(t-\tau),
    \label{eq:phase_kernel}
\end{equation}
where \(K_0\) and \(K_\tau\) denote the instantaneous and delayed coupling
strengths, respectively, and \(\phi\) is the propagation phase between the two coupling points.
Varying \(\phi\) leaves the return time \(\tau\) unchanged,
but changes the phase with which the delayed field re-enters the resonator. It therefore controls the coherent interference between the returning field
and the instantaneous resonator response.

\begin{figure}[t]
    \centering
    \includegraphics[width=0.85\columnwidth]{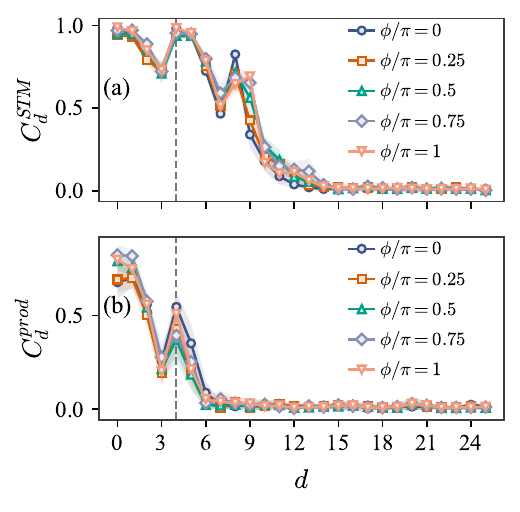}
    \caption{
    Propagation-phase dependence of delay-resolved memory diagnostics at fixed
    programmed delay \(\tau/\Delta t=4\).
    (a) STM capacity \(C_d^{\rm STM}\).
    (b) Delayed-product capacity \(C_d^{\rm prod}\).
    Different curves correspond to
    \(\phi/\pi\in\{0,0.25,0.5,0.75,1\}\).
    The vertical dashed line marks \(d=\tau/\Delta t\).
    }
    \label{fig:app_phase_scan}
\end{figure}

We scan \(\phi/\pi\in\{0,0.25,0.5,0.75,1\}\) at fixed delay
\(\tau/\Delta t=4\), using the same TDQRC and readout protocol as in Fig.~\ref{fig:memory_product}. The resulting STM and delayed-product capacities are shown in Fig.~\ref{fig:app_phase_scan}. Obviously, the propagation phase primarily controls the strength and shape of the memory revival, while leaving its temporal position approximately unchanged.
\label{app:phase_scan}
In both the STM and delayed-product diagnostics, the feedback-induced feature remains localized near
\(d=\tau/\Delta t\), because its arrival time is fixed by the propagation delay. Its amplitude and surrounding lag profile, however, vary with \(\phi\) owing to constructive or destructive interference between the
returning field and the resonator response. The choice \(\phi=0\) used in the main text therefore represents the constructive-feedback configuration in
which the delay-induced memory feature is most clearly resolved.

\section{Feedback-amplitude tuning}

The geometry-programmed memory kernel provides separate control over the
arrival time, phase, and weight of the coherently returned field. The
propagation delay \(\tau\) determines when previously emitted information
returns to the resonator, while the coupling factor
\(\sqrt{\gamma_\mu\gamma_\nu}\) determines the weight of the corresponding
delayed-feedback contribution.
To isolate the latter effect, we vary
\(\sqrt{\gamma_1\gamma_2}/\Gamma\) while keeping the total coupling rate
\(\Gamma=\gamma_1+\gamma_2\) and the programmed delay
\(\tau/\Delta t\) fixed. The accessible range is
\begin{equation}
    0
    \leq
    \frac{\sqrt{\gamma_1\gamma_2}}{\Gamma}
    \leq
    \frac{1}{2},
    \label{eq:feedback_amplitude_range}
\end{equation}
where the upper bound is reached for balanced couplings,
\(\gamma_1=\gamma_2\).

\begin{figure}[t]
    \centering
    \includegraphics[width=0.85\columnwidth]{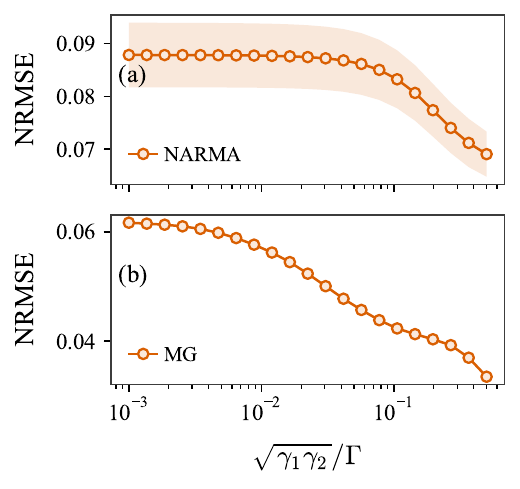}
    \caption{
    Prediction performance as a function of the normalized delayed-feedback coupling factor at fixed
    \(\Gamma=\gamma_1+\gamma_2=0.2\), \(U=6\), and
    \(\tau/\Delta t=4\).
    The horizontal axis is \(\sqrt{\gamma_1\gamma_2}/\Gamma\).
    (a) Test NRMSE for the NARMA-5 task.
    (b) Forecasting NRMSE for the Mackey-Glass task described in Appendix~\ref{app:mg}, evaluated at prediction horizon \(D=10\).
    }
    \label{fig:feedback_amplitude}
\end{figure}

As the delayed-feedback amplitude increases, both prediction tasks improve over the parameter range considered [Fig.~\ref{fig:feedback_amplitude}]. At fixed total coupling rate and propagation delay, a larger
value of \(\sqrt{\gamma_1\gamma_2}\) increases the contribution of the coherently returned field to the reservoir dynamics, thereby strengthening the task-relevant delayed temporal features.
The balanced  point \(\gamma_1=\gamma_2\), which maximizes
\(\sqrt{\gamma_1\gamma_2}\)  at fixed total decay rate, 
therefore realizes the strongest coherent-feedback regime and is adopted in the main-text
simulations.

Apart from the propagation delay and phase discussed in the previous
appendix, the delayed-feedback amplitude thus provides complementary parameters
for engineering the geometry-programmed memory kernel.
\label{app:gamma_scan}

\section{Feature-Level initial-state washout}
\label{app:esp}

The delay-induced memory revivals discussed in the main text should encode the history of the driven input sequence rather than a persistent dependence on the initial resonator state. We therefore examine whether two reservoirs subjected
to the same input sequence but initialized differently develop convergent readout features. This provides an empirical feature-level check of
initial-state forgetting, as required for echo-state behavior.

We initialize the resonator in
\begin{equation}
    |\psi_0^{(1)}\rangle=|0\rangle,
    \qquad
    |\psi_0^{(2)}\rangle=|1\rangle,
\end{equation}
while taking the waveguide field to be initially in vacuum in both cases. The two reservoirs are then driven by the same input sequence using otherwise identical parameters.
Let \(\mathbf x_k^{(1)}\) and \(\mathbf x_k^{(2)}\) denote the corresponding
readout-feature vectors at input step \(k\), with the constant bias component excluded. We quantify their separation through
the normalized Euclidean
distance
\begin{equation}
    D_k
    =
    \frac{
    \|\mathbf x_k^{(1)}-\mathbf x_k^{(2)}\|_2
    }{
    \sqrt{
    \left(
    \|\mathbf x_k^{(1)}\|_2^2+
    \|\mathbf x_k^{(2)}\|_2^2
    \right)/2
    }
    +\eta
    },
    \label{eq:esp_distance}
\end{equation}
where \(\eta=10^{-12}\) prevents numerical singularities when both feature
norms become very small.

\begin{figure}[t]
    \centering
    \includegraphics[width=0.85\columnwidth]{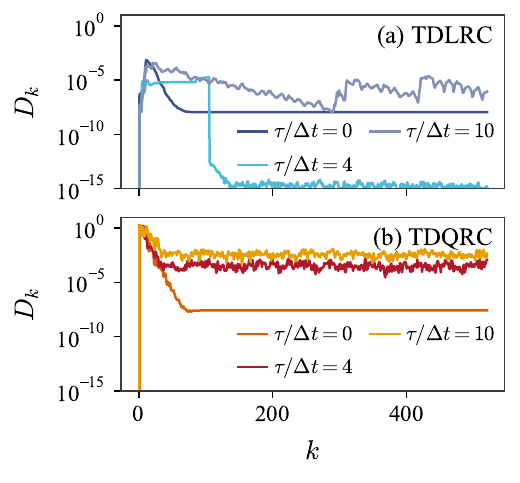}
    \caption{
    Feature-level initial-state washout for reservoirs driven by the same
    input sequence but initialized in the resonator states \(\ket{0}\) and
    \(\ket{1}\).
    The normalized feature distance \(D_k\), defined in
    Eq.~\eqref{eq:esp_distance}, is shown for the linear and Kerr-nonlinear reservoirs at representative delays \(\tau/\Delta t\in\{0,4,10\}\). 
    }
    \label{fig:esp}
\end{figure}

The feature distance $D_k$ decreases by several orders of magnitude during the washout interval[Fig.~\ref{fig:esp}]. Consequently, the readout features used for training and testing become insensitive to
the chosen initial resonator state.
 This supports the interpretation that the delay-resolved revivals reported in the main text represent driven input-history memory rather than an unwashed initial-state contribution. The Kerr-nonlinear reservoir exhibits a larger long-time feature mismatch than the linear reference for the parameters considered. This indicates slower or less complete feature-level convergence in the nonlinear dynamics, although the remaining difference is small relative to the initial separation.
For the linear reservoir at long propagation delay, weak bounded oscillations remain
visible in \(D_k\). Their characteristic delayed recurrence is consistent with
coherent returns of the residual initial-state component through the feedback
line.
\\

\section{Non-integer programmed delays}
\label{app:noninteger-delay}

The main text considers integer delay ratios $d_r=\tau/\Delta t$,  for which the propagation time can be associated directly with an integer input lag. 
In a waveguide implementation, however, \(\tau\) is determined by the propagation length and group velocity and need not be commensurate with the input-update interval \(\Delta t\). We therefore examine half-integer values of \(d_r\) to test whether the delay-controlled memory revival persists away from this commensurate limit.

The benchmark targets are defined only at integer lags $d$. Letting $d_r=m+\eta$ with $m=\lfloor d_r\rfloor$ and $0<\eta<1$, the returning field overlaps two adjacent input bins,
\begin{equation}
    u(t-\tau)\;\longrightarrow\;(1-\eta)\,u_{k-m}+\eta\,u_{k-m-1},
\end{equation}
up to the sampling convention and the temporal filtering of the resonator and virtual-node readout. A fractional delay thus produces no target at a non-integer lag; its weight is projected onto $d=m$ and $d=m+1$, broadening or splitting the revival.

\begin{figure}[t]
    \centering
    \includegraphics[width=\columnwidth]{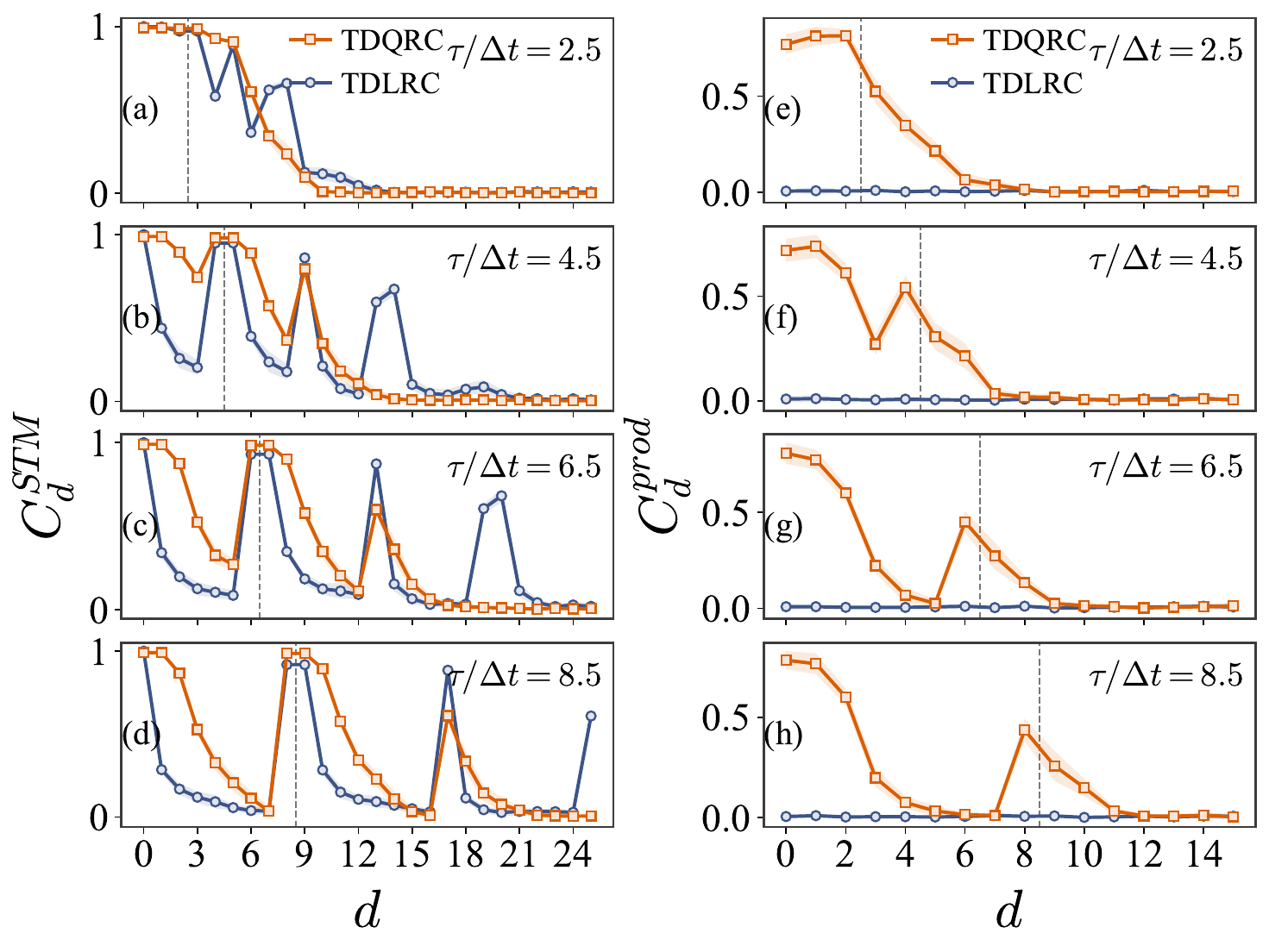}
    \caption{ Delay-resolved memory capacities for noninteger programmed delay ratios.
    (a)--(d) Short-term-memory capacity $C_d^{\rm STM}$ for $\tau/\Delta t=2.5,4.5,6.5,8.5$. (e)--(h) Corresponding delayed-product capacity $C_d^{\rm prod}$.  The vertical dashed line indicates the programmed return time
    $d_r=\tau/\Delta t$. Because the benchmark targets are defined at
    integer lags, a noninteger return time is resolved as a distributed
    response over the neighboring values of \(d\). The Kerr-nonlinear TDQRC
    exhibits a corresponding delayed-product feature, whereas the linear
    reference has negligible product capacity.}
    \label{fig:noninteger-delay}
\end{figure}

The resulting STM and delayed-product capacities for $\tau/\Delta t=2.5,,4.5,,6.5,$ and $8.5$ are shown in Fig.~\ref{fig:noninteger-delay}. The revivals in $C_d^{\rm STM}$ continue to track the programmed delay.  
For the longer delays, the primary revival is shared by the two integer lags adjacent to \(d_r\), while weaker features at later lags arise from repeated coherent returns.
For $d_r=2.5$, the returned field overlaps more strongly with the intrinsic
short-time response of the reservoir, producing a broader low-lag structure
rather than a clearly separated revival. 

The delayed-product capacity exhibits the corresponding nonlinear response.
The Kerr interaction mixes the coherently returned field with the current
reservoir dynamics, generating delayed nonlinear features associated with both
neighboring integer lags. The resulting product-capacity profile therefore
develops a broadened shoulder or a pair of neighboring features around
\(d_r\). This structure is most clearly resolved for the longer propagation
delays, where it is separated from the intrinsic short-time nonlinear
response. By contrast, the linear reference reservoir exhibits negligible
delayed-product capacity over the parameter range shown, confirming that
coherent delayed memory alone does not generate the required multiplicative
feature under a linear readout.

These results extend the delay-matching picture of the main text beyond
integer delay ratios. When \(\tau/\Delta t\) is an integer, the coherent
return is localized near a single discrete lag. When it is noninteger, the
same return time is represented by a distributed response over neighboring
integer-lag targets. Thus the propagation delay continues to determine the
temporal location of the memory feature, while the discrete sampling protocol
determines how that feature is resolved by the readout.

\section{Multi-delay geometry}
\label{app:multi_delay}

The two-point geometry considered in the main text is the minimal configuration
that produces a single coherent memory return and therefore allows the roles of
memory placement and Kerr-induced nonlinear processing to be distinguished
most clearly.
Adding further coupling points naturally extends the
geometry-programmed memory kernel by introducing several delayed feedback
paths. Their computational benefit, however, is not automatic, because the
corresponding coherently returned fields can interfere before being processed
by the Kerr nonlinearity.
For a three-point geometry, the kernel takes the form
\begin{equation}
    K(t)
    \simeq
    K_0\delta(t)
    +
    K_{12}\delta(t-\tau_{12})
    +
    K_{23}\delta(t-\tau_{23})
    +
    K_{13}\delta(t-\tau_{13}),
    \label{eq:three_point_kernel}
\end{equation}
where \(\tau_{\mu\nu}=|x_\mu-x_\nu|/v_g\), while the coefficients
\(K_{\mu\nu}\) include the corresponding coupling strengths and propagation
phases.

\begin{figure}[t]
    \centering
    \includegraphics[width=0.85\columnwidth]{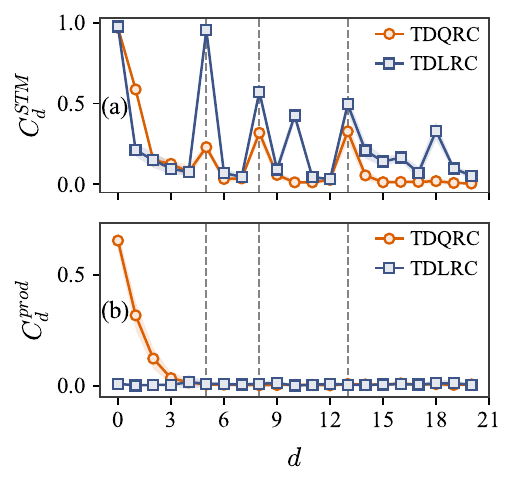}
    \caption{Delay-resolved memory diagnostics for a three-point geometry.
    (a) STM capacity \(C_d^{\rm STM}\), showing revivals near the programmed
    propagation delays.
    (b) Delayed-product capacity \(C_d^{\rm prod}\). Vertical dashed lines
    indicate the programmed delay indices.
    }
    \label{fig:app_three_delay}
\end{figure}

Distinct STM revivals appear near the programmed delays in Fig.~\ref{fig:app_three_delay}(a), confirming that the placement of linear memory generalizes naturally to geometries with multiple feedback paths. Each delayed component
of the kernel produces a corresponding coherent memory return, although the
observed amplitudes and line shapes also depend on interference between the
different channels. The delayed-product response in
Fig.~\ref{fig:app_three_delay}(b) exhibits qualitatively different behavior. Rather than producing one pronounced nonlinear-memory peak at each programmed
delay, the TDQRC displays mainly a short-time nonlinear response in the
parameter regime considered, with no clearly resolved delayed-product revival
at the individual return times. The TDLRC remains close to the product-capacity
floor, as expected for a linear reservoir with a linear readout. These results
show that adding coupling points directly programs several linear-memory return
times, but the associated nonlinear processing is not determined by the delay
positions alone. The Kerr interaction acts on the superposition of the
instantaneous response and several coherently returned fields, so the resulting
nonlinear features depend jointly on the feedback weights, propagation phases,
temporal overlap, and interference between delayed channels. The two-point
geometry used in the main text therefore provides the minimal setting in which
coherent memory placement and nonlinear memory conversion can be separated
unambiguously. More generally, multipoint geometries offer a route toward
hardware-programmable memory kernels with several temporal scales, which may
be useful for tasks involving distributed temporal dependencies once the
relative phases and feedback amplitudes are optimized.

\section{Product-capacity diagnostic for NARMA task matching}
\label{app:narma_product_matching}

\begin{figure}[t]
    \centering
    \includegraphics[width=\linewidth]{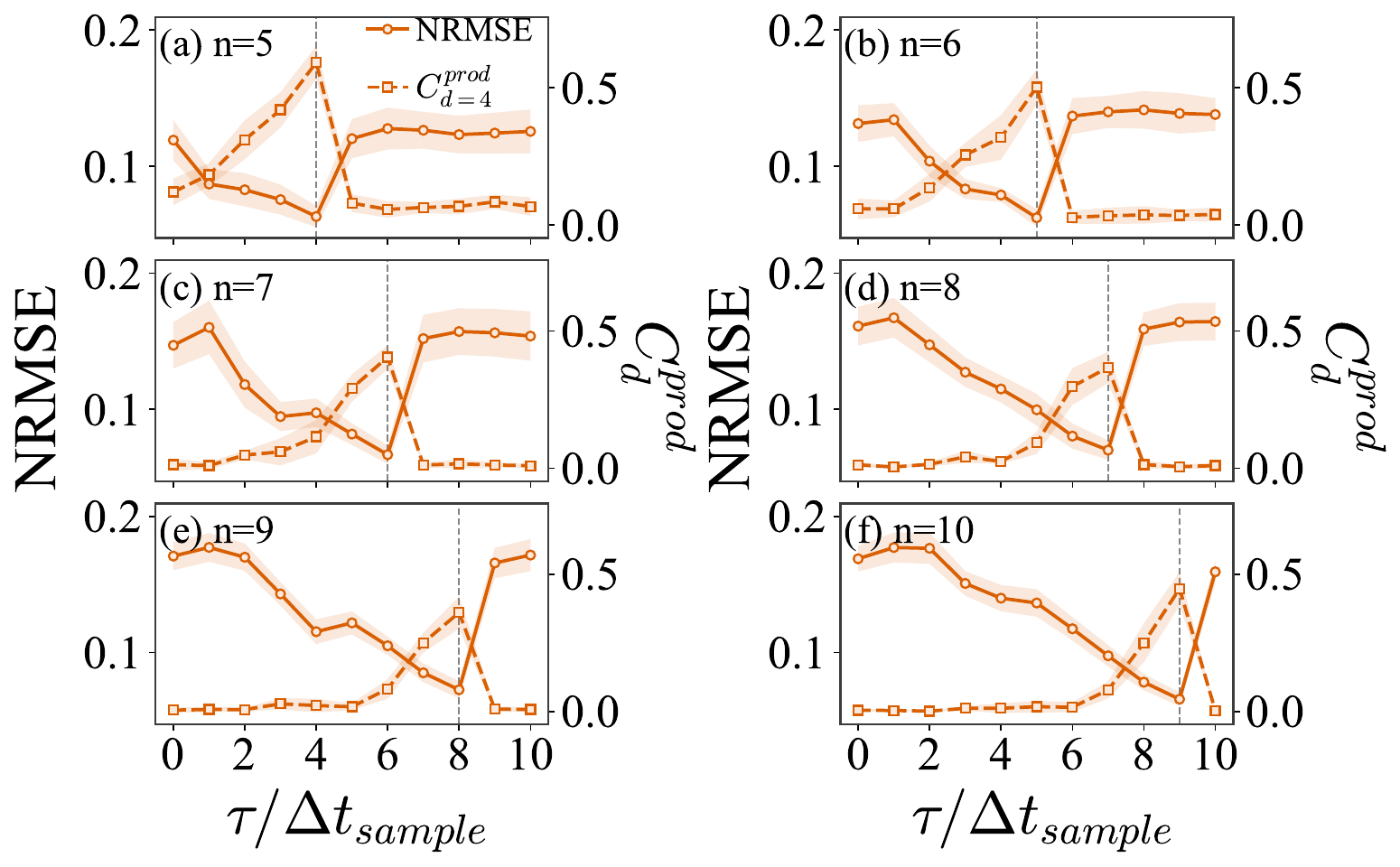}
    \caption{
    Comparison between NARMA-$n$  prediction error and the delayed-product
    capacity evaluated at the task-relevant lag \(d=n-1\).
    Solid circles (left axis) show the test NRMSE. 
    Dashed squares (right axis) show $C^{\rm prod}_{d=n-1}$.
    The vertical dashed line marks the task-matched delay $\tau/\Delta t=n-1$. Shaded regions  indicate one standard deviation over independent input realizations.    }
    \label{fig:app_narma_product_matching}
\end{figure}

The main text argues that the optimal programmed delay is determined by the
explicit nonlinear temporal dependence of the target task. For the NARMA-$n$ benchmark defined in Eq.~\eqref{eq:narma},   the only explicit input--input
nonlinearity is
\begin{equation}
    1.5\,\tilde u_k \tilde u_{k-n+1},
\end{equation}
which defines the task-relevant product lag
\begin{equation}
    d_{\rm task}=n-1.
\end{equation}
To exame whether the prediction optimum   associated with this lag, 
we compare the NARMA-$n$ prediction error with the delayed-product capacity
$C^{\rm prod}_{d=n-1}$, evaluated independently using the diagnostic task of
Sec.~\ref{sec:memory}, with  the same reservoir parameters and linear readout protocol.

Figure~\ref{fig:app_narma_product_matching} provides an independent mechanistic check of the matching picture. As the task order increases
from $n=5,\ldots,10$, the minimum prediction error shifts from \(\tau/\Delta t\simeq4\) to \(\tau/\Delta t\simeq9\). Over the same parameter
range, the delayed-product capacity at the corresponding task lag,
$C^{\rm prod}_{d=n-1}$, develops its maximum at nearly the same programmed
delay. Thus, lower prediction error coincides with stronger nonlinear delayed
features at the explicit NARMA product lag. The observed correlation is
consistent with the interpretation that the optimal programmed delay is
governed by the availability of task-relevant nonlinear memory, rather than by
memory depth alone.

\section{Semiclassical nonlinear-resonator control}
\label{app:crc}

To determine whether the delayed nonlinear features of the TDQRC can be
reproduced by a classical mean-field oscillator, we compare it with a
semiclassical nonlinear-resonator control (CRC). 
Starting from the operator equation of motion, we replace the resonator operator by a complex amplitude,
\(a(t)\rightarrow\alpha(t)\) and and close the nonlinear term through the
mean-field factorization
\(a^\dagger a a\rightarrow |\alpha|^2\alpha\). 
For the two-point geometry with constructive feedback, the resulting
delay-differential equation is
\begin{equation}
    \dot\alpha(t)
    =
    -\frac{\Gamma}{2}\alpha(t)
    -
    \frac{\sqrt{\gamma_1\gamma_2}}{2}\alpha(t-\tau)
    -
    iU|\alpha(t)|^2\alpha(t)
    +
    \Omega(t),
    \label{eq:crc_two_point}
\end{equation}
using the same kernel convention as in Eq.~\eqref{eq:geometry_kernel}. 
The CRC therefore retains both the programmed delayed-feedback channel and
Kerr-type
nonlinearity, while reducing the joint resonator--waveguide quantum dynamics to a single deterministic
complex trajectory.
In particular, it neglects quantum fluctuations,
higher-order moment dynamics, and correlations between the resonator and the
propagating field.

\begin{figure}[t]
    \centering
    \includegraphics[width=\linewidth]{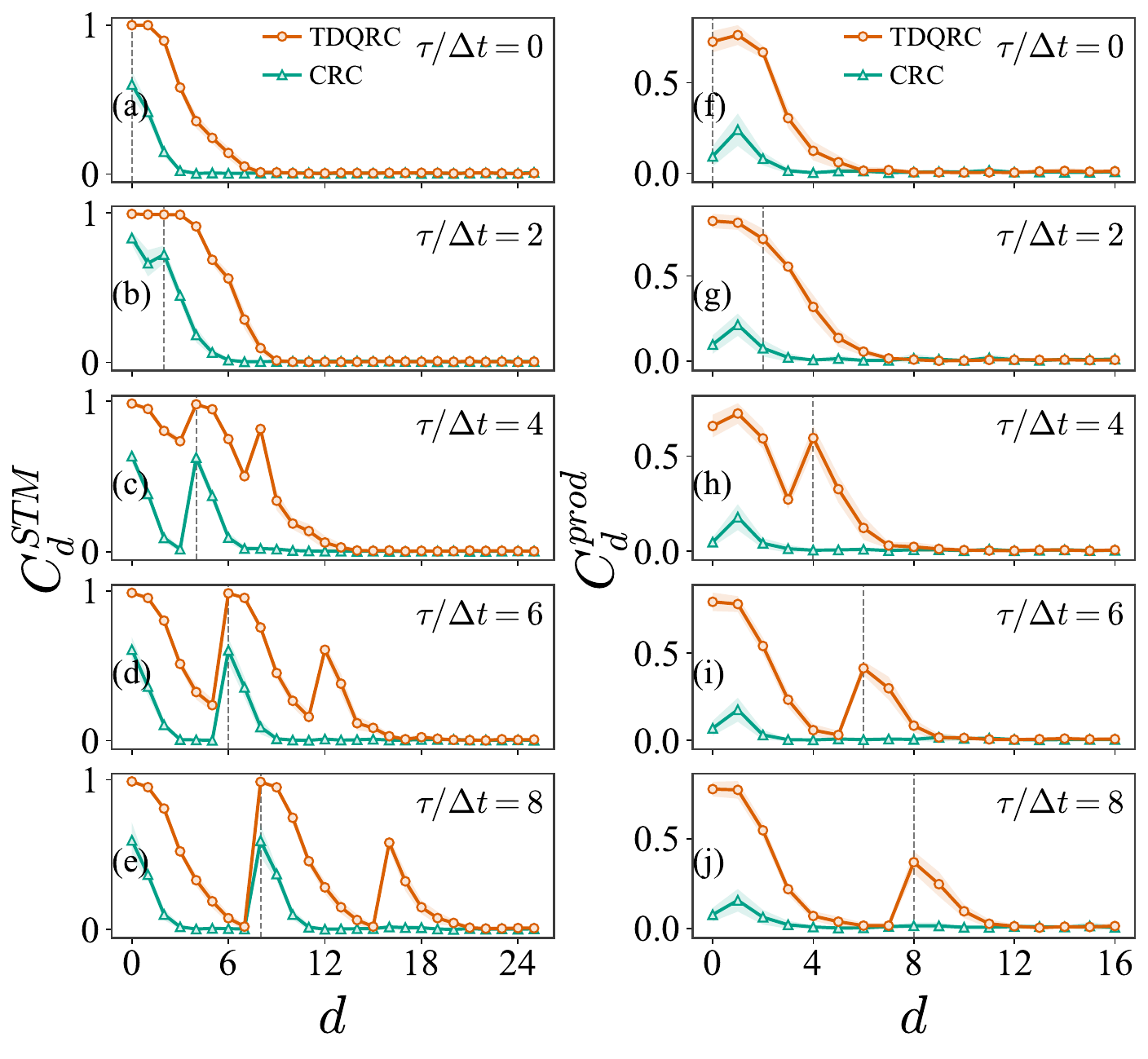}
    \caption{
      Comparison of the TDQRC with the semiclassical nonlinear-resonator control  for delay-resolved memory diagnostics. 
    Left column shows the STM capacity \(C_d^{\rm STM}\) and the right column shows the  delayed-product capacity \(C_d^{\rm prod}\). Rows correspond to different  programmed delays \(\tau/\Delta t\). 
    Vertical dashed lines mark  the nominal coherent-return lag \(d=\tau/\Delta t\).
    }
    \label{fig:app_crc_memory}
\end{figure}
\begin{figure}[t]
    \centering
    \includegraphics[width=\linewidth]{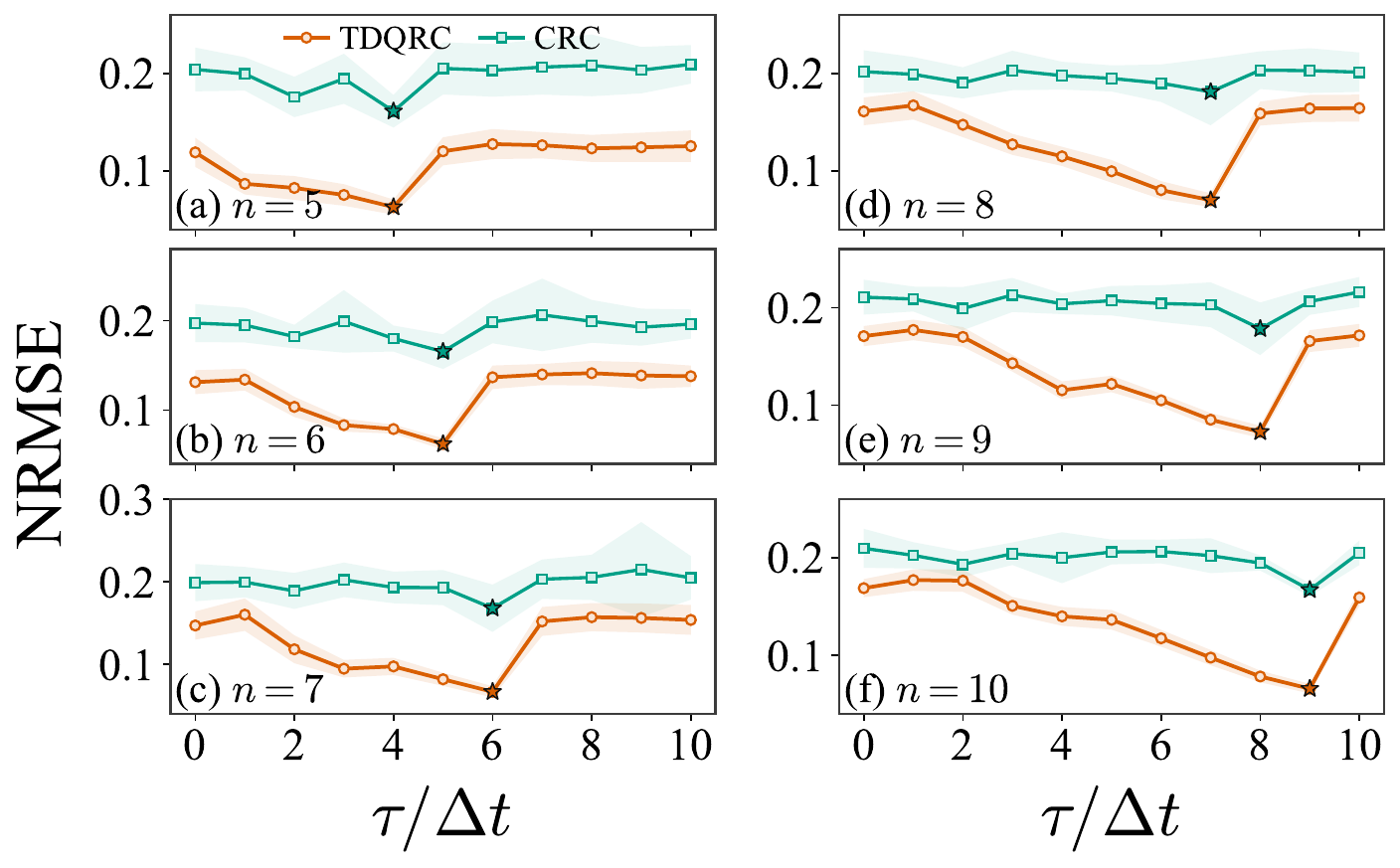}
    \caption{   Comparison of the TDQRC and semiclassical nonlinear-resonator control for  NARMA prediction. 
    The test NRMSE is plotted as a function of programmed delay 
    \(\tau/\Delta t\) for
    \(n=5,\ldots,10\). 
    Stars indicate the minimum NRMSE attained by each curve.
    }
    \label{fig:app_crc_narma}
\end{figure}

The CRC readout is constructed from the classical resonator quadratures
\begin{equation}
    X_\alpha(t)=\alpha(t)+\alpha^*(t),
    \qquad
    P_\alpha(t)=i[\alpha^*(t)-\alpha(t)],
\end{equation}
together with the corresponding output-field quadratures. 
The same input
encoding, virtual-node sampling, feature normalization, and linear readout
protocol are used for the CRC and TDQRC, so that the comparison isolates the
difference between deterministic mean-field dynamics and the full quantum
evolution.

Figure~\ref{fig:app_crc_memory} shows that the CRC reproduces the primary STM
revival near the programmed delay. This confirms that the temporal placement
of linear memory is already captured by the classical delayed-feedback
equation and is primarily determined by waveguide propagation. A clearer
difference emerges in the delayed-product diagnostic. Over the parameter range
shown, the CRC develops substantially weaker nonlinear delayed-memory capacity
than the TDQRC and does not reproduce the full delay-dependent product-feature
profile. The programmed delay alone is therefore insufficient to account for
the nonlinear memory observed in the quantum reservoir, while a single
mean-field Kerr trajectory also provides only a partial description of its
nonlinear processing.
The NARMA results in Fig.~\ref{fig:app_crc_narma} show the corresponding
difference at the task level. Although the CRC benefits from the same
geometry-programmed feedback delay, its prediction error remains higher than
that of the TDQRC for the configurations considered. Together with the
delayed-product diagnostic, this result shows that the useful nonlinear
temporal features of the TDQRC are not fully reproduced by deterministic
scalar mean-field dynamics. The comparison therefore identifies a
beyond-mean-field contribution to the reservoir response, arising from quantum
fluctuations, higher-order correlations, or joint resonator--field dynamics
that are absent from the CRC.

\section{Mackey-Glass forecasting}
\label{app:mg}

\begin{figure}[t]
    \centering
    \includegraphics[width=\linewidth]{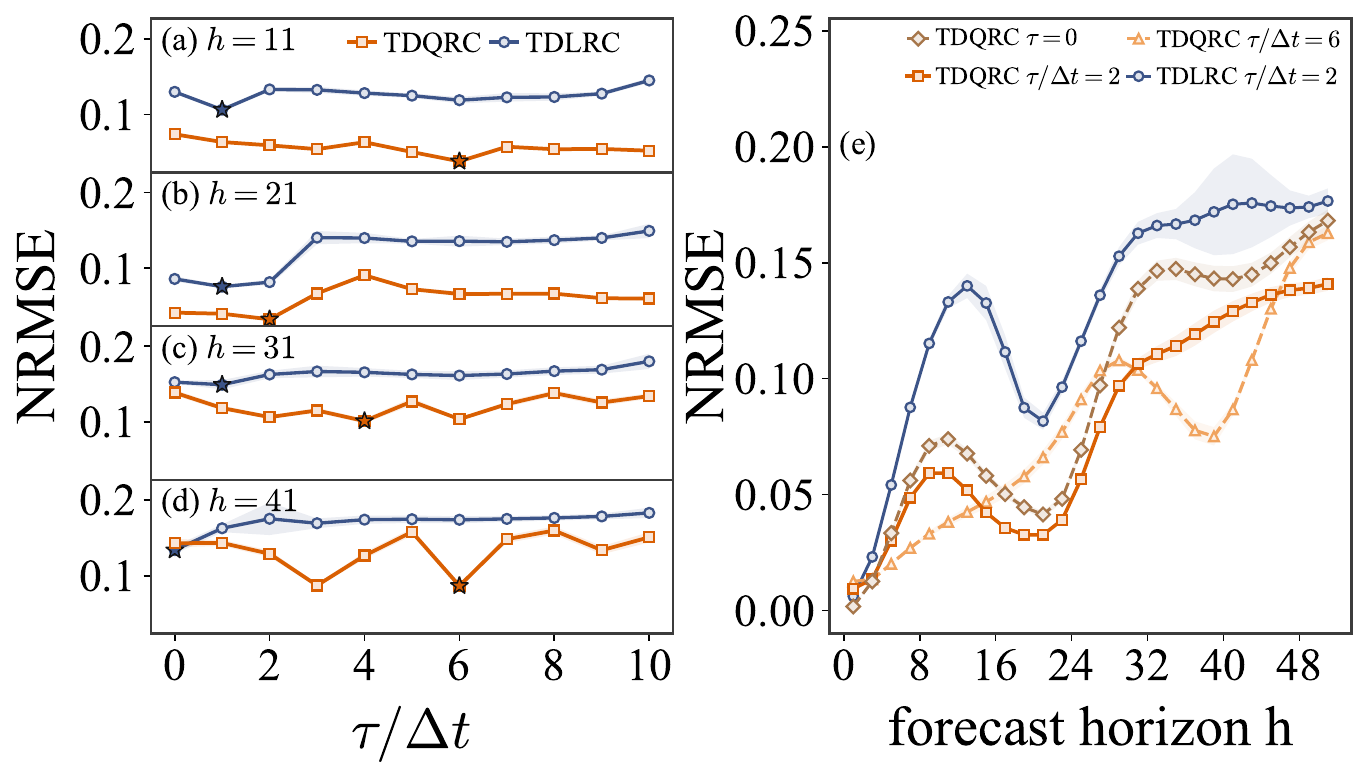}
    \caption{
    Mackey--Glass forecasting with programmable delayed feedback.
    (a)--(d) Test NRMSE versus programmed delay \(\tau/\Delta t\) for forecast
    horizons \(h\in\{11,21,31,41\}\), comparing the TDQRC and TDLRC. 
    Stars indicate  the minimum NRMSE within the scanned delay range.
    (e) Test NRMSE versus forecast horizon for representative programmed
    delays.
    The case \(\tau/\Delta t=0\) corresponds to  the zero-delay
    limit.
    }
    \label{fig:app_mg}
\end{figure}

To examine whether the task-matching picture extends beyond synthetic
delay benchmarks, we consider forecasting of the chaotic
Mackey--Glass time series~\cite{vrugt2026}.
The continuous Mackey--Glass signal \(s(t)\) is generated from
\begin{equation}
    \frac{d s(t)}{dt}
    =
    \beta_{\rm MG}
    \frac{s(t-T_{\rm MG})}
    {1+s(t-T_{\rm MG})^{p_{\rm MG}}}
    -
    \gamma_{\rm MG}s(t),
    \label{eq:mg_dde}
\end{equation}
with \(\beta_{\rm MG}=0.2\), \(\gamma_{\rm MG}=0.1\),
\(p_{\rm MG}=10\), and \(T_{\rm MG}=17\). The signal is sampled with
\(\delta t_{\rm MG}=0.1\) to fobtain
\(x_k=s(k\delta t_{\rm MG})\). For forecast horizon \(h\), the predication target is
\(y_k^{(h)}=x_{k+h}\). 
The input encoding, virtual-node construction,
washout procedure, training/test partition, and linear readout are the same as those used throughout the main text.

Unlike the NARMA-\(n\) benchmarks, the Mackey--Glass task does not possess a
single dominant delayed-product feature. Instead, its predictive information
is distributed over multiple temporal correlations generated by the underlying
nonlinear delay dynamics. Consequently, no unique task-relevant lag satisfying
\(\tau/\Delta t=d_{\rm task}\) is expected. Fig.~\ref{fig:app_mg} is consistent with this picture. The TDQRC generally
outperforms the TDLRC over the forecast horizons considered, indicating that
Kerr-induced nonlinear processing remains useful for chaotic time-series
prediction. In contrast to the sharper NARMA behavior, however, the dependence
on the programmed delay is broad and horizon dependent. Different forecast
horizons favor different delay values, and increasing the delay alone does not
systematically improve the forecasting performance.

The delay-matching principle therefore applies most naturally to tasks with a
dominant temporal scale, whereas for tasks with distributed temporal
correlations the programmed delay should be regarded as a tunable memory
timescale rather than a uniquely determined matching condition.

\bibliography{qrc}

\end{document}